\documentclass{PoS}

\usepackage{microtype}
\usepackage[english]{babel}
\usepackage[utf8]{inputenx}	
\usepackage{cite}
\usepackage{graphicx}
\usepackage{amsmath}
\usepackage{mathtools}
\usepackage{amsthm}
\usepackage{subfig}
\pdfoutput=1

\title{New extended interpolating operators for hadron correlation functions}

\ShortTitle{New extended interpolating operators for hadron correlation functions}

\author{\speaker{Francesco Scardino}\\
        Dipartimento di Fisica, "Sapienza" Università di Roma, and INFN, Sezione di Roma, Piazzale Aldo Moro 2, I-00185 Roma, ITALY. \\
        E-mail: \email{francesco.scardino@roma1.infn.it}}

\author{Mauro Papinutto\\
       Dipartimento di Fisica, "Sapienza" Università di Roma, and INFN, Sezione di Roma, Piazzale Aldo Moro 2, I-00185 Roma, ITALY. \\
       E-mail: \email{mauro.papinutto@roma1.infn.it}}
    
    \author{Stefan Schaefer\\
    	Neumann Institute for Computing, DESY, Platanenallee 6, 15738 Zeuthen, Germany\\
    	E-mail: \email{stefan.schaefer@desy.de}}

\abstract{New extended interpolating operators made of quenched three
dimensional fermions are introduced in the context of lattice QCD. The mass of
the 3D fermions can be tuned in a controlled way to find a better overlap of
the extended operators with the states of interest. The extended operators have
good renormalisation properties and are easy to control when taking the
continuum limit. Moreover the short distance behaviour of the two point
functions built from these operators is greatly improved.	The operators have
been numerically implemented and a comparison to point sources and Jacobi
smeared sources has been performed on the new CLS configurations.\\
\mbox{}\hfill{\tt DESY-16-247 }
}

\FullConference{34th annual International Symposium on Lattice Field Theory\\
		24-30 July 2016\\
		University of Southampton, UK}

\begin{document}

\section{Introduction}

A serious problem which limits the attainable precision in lattice QCD
computations of hadronic observables is the exponential suppression of the
signal-to-noise ratio in Euclidean time. It is therefore important to reduce
the coefficient of this deterioration as much as possible.

One first step is to find improved interpolating
operators, which allow to decrease statistical and systematic errors in the
extraction of the hadron spectrum, form factors and matrix elements. For this
purpose, we introduce  extended interpolating operators built from 3D
fermions. These are well behaved under renormalisation and improve the short
distance behaviour of two point functions. The octet and decuplet baryonic
interpolating operators have been classified according to the irreducible
representations of the cubic group \cite{Basak:2005aq}\cite{Basak:2005ir}. This
classification has the advantage of being exhaustive by construction.

In this proceedings contribution we first review the construction of the 
basis of baryon operators, before introducing the new source operators.
They are then compared to the traditional Jacobi smearing.

\section{Classification of baryonic operators}
Baryons are bound states of three valence quarks. The interpolating operators $\mathcal{B}$ will need to have a well defined set of quantum numbers such that the corresponding Hilbert space operator $\hat{\mathcal{B}}$ projects onto the state we are interested in.
 On the lattice, rotational symmetry breaks down and is replaced by the cubic group $SO(3,\mathbb{Z})$  whose elements are the matrices of $SO(3)$ with integer entries. In order to correctly classify the operators, we will use the spin covering group  $Spin(3,\mathbb{Z})$ of  $SO(3,\mathbb{Z})$. In fact the $Spin(3)$ group is isomorphic to $SU(2)$ and  allows us to work with the continuum Weyl spinor notation on the lattice.\par
 A generic gauge invariant three-quark operator will have the form
 \begin{equation}
 \mathcal{B}(x) = u^{a}_{\alpha}(x)d^{b}_{\beta}(x)s^{c}_{\gamma}(x)\ t^{\alpha\beta\gamma}\epsilon_{abc},
 \end{equation}
 where $u,d,s$ are the quark fields, which are in the fundamental representation of $SU(3)_c$. Furthermore, despite the notation we use, the fields have a definite but not necessarily different flavour. The tensor $t^{\alpha \beta \gamma}$ depends on the $SO(3,\mathbb{Z})$ representation the baryon falls in,  $\epsilon_{abc}$ is the only allowed invariant colour tensor. Greek indices $\alpha,\beta,\ldots$ are Dirac indices while the Latin ones $(a,b,\ldots)$ represent colour. \par
In order to classify the baryonic operators according to the irreducible representations of flavour and the rotation group on the lattice, we need  to classify the tensors $t^{\alpha\beta\gamma}$ introduced above according to the irreducible representations of the spin covering $Spin(3,\mathbb{Z})$ of the cubic group $SO(3,\mathbb{Z})$. For spin $\frac{1}{2}$ and spin $\frac{3}{2}$ baryons, it can be proven \cite{Johnson:1982yq} that there is a one to one correspondence between the corresponding $Spin(3,\mathbb{Z})$ representations on the lattice and the continuum ones. We thus use the continuum notation of dotted and undotted Weyl spinors in the following.\par
 We now discuss the operators that have been actually used in the simulations. Consider the case in which two flavours are equal, namely for spin $s=\frac{1}{2}$ the case of the nucleon. It is possible to show that there are two proton operators $\hat{p}$ and $\tilde{p}$ in the $\left(\frac{1}{2},0\right)$ representation 
 \begin{align}
 \label{proton12}
 \nonumber
 &\hat{p}^{(\frac{1}{2},0)}_{\frac{1}{2},\frac{1}{2}}=\frac{1}{\sqrt{2}}(u_1 d_2-u_2 d_1)u_1\qquad\qquad \hat{p}^{(\frac{1}{2},0)}_{\frac{1}{2},-\frac{1}{2}}=\frac{1}{\sqrt{2}}(u_1 d_2-u_2d_1)u_2\\
 &\tilde{p}^{(\frac{1}{2},0)}_{\frac{1}{2},\frac{1}{2}}=\frac{1}{\sqrt{2}}(u_{\dot{1}}d_{\dot{2}}-u_{\dot{2}}d_{\dot{1}})u_1\qquad\qquad \tilde{p}^{(\frac{1}{2},0)}_{\frac{1}{2},-\frac{1}{2}}=\frac{1}{\sqrt{2}}(u_{\dot{1}}d_{\dot{2}}-u_{\dot{2}}d_{\dot{1}})u_2
 \end{align}
 and one in the $\left(\frac{1}{2},1\right)$ representation 
 \begin{align}
 \label{proton3}
 \nonumber
 &{p}^{(\frac{1}{2},1)}_{\frac{1}{2},\frac{1}{2}}=\frac{1}{\sqrt{14}}\{(u_{\dot{1}}d_{\dot{2}}+u_{\dot{2}}d_{\dot{1}})u_1-2u_{\dot{1}}d_{\dot{1}}u_2+2(u_{\dot{1}}d_2-u_{\dot{2}}d_1)u_{\dot{1}}\}\\ &{p}^{(\frac{1}{2},1)}_{\frac{1}{2},-\frac{1}{2}}=-\frac{1}{\sqrt{14}}\{(u_{\dot{1}}d_{\dot{2}}+u_{\dot{2}}d_{\dot{1}})u_2-2u_{\dot{2}}d_{\dot{2}}u_1+2(u_{\dot{2}}d_1-u_{\dot{1}}d_2)u_{\dot{2}}\}.
 \end{align}
In the case of three equal flavours, namely the case of the $\Omega$ baryon, there are two $s = \frac{3}{2}$ decouplet operators
 \begin{align}
 \label{omega12}
 \nonumber
 &{\Omega}^{(\frac{1}{2},1)}_{\frac{3}{2},\frac{3}{2}}=s_{\dot{1}}s_{\dot{1}}s_1 \qquad\qquad {\Omega}^{(\frac{1}{2},1)}_{\frac{3}{2},\frac{1}{2}}=\frac{1}{\sqrt{3}}\{s_{\dot{1}}s_{\dot{1}}s_2+s_{\dot{1}}s_{\dot{2}}s_1+s_{\dot{2}}s_{\dot{1}}s_1 \}\\
 \nonumber
 &{\Omega}^{(\frac{3}{2},0)}_{\frac{3}{2},\frac{3}{2}}=s_1s_1s_1 \qquad\qquad {\Omega}^{(\frac{3}{2},0)}_{\frac{3}{2},\frac{1}{2}}=\frac{1}{\sqrt{3}}\{s_1s_1s_2+s_1s_2s_1+s_2s_1s_1 \}\,,\\
 \end{align}
 where we have reported only the spin up components. In order to have operators with definite transformation properties under the action of parity $\hat{\mathcal{P}}$, we need to consider
 \begin{align}
 \label{parityops}
 &\mathcal{B}^{\pm}_{(a,b)\oplus(b,a)} =\mathcal{B}_{(a,b)}\mp\mathcal{B}_{(b,a)} \ ,
 \end{align}
 where $\pm$ indicates the parity eigenvalue and $(a,b)$ the irreducible representation. 
\section{Extended operators}
The 3D extended operators are built from quenched three dimensional fermions fields coupled via pseudoscalar bilinears with ordinary four dimensional fermions in the bulk.
These 3D fields live on a time-slice and their propagator can be derived from the action
\begin{equation}
\label{3daction}
S^{3D} = a^3 \sum_{\mathbf{x}} \bar{\varphi}(\mathbf{x})\mathcal{D}\varphi(\mathbf{x}), \qquad \mathcal{D} = \frac{1}{2}\sum_{i=1}^{3}\{\gamma_i(\nabla^{*}_i+\nabla_i)-a\nabla^{*}_i\nabla_i\}+{m_{3D}}
\end{equation}
where $\mathcal{D}$ is the three dimensional Wilson-Dirac operator with a mass
term $m_{3D}$. The 3D fermionic fields
$\varphi$,$\bar{\varphi}$ are spin $1/2$ spinors with canonical dimension $1$
and represent a flavour $SU(3)$ triplet $\bar{\varphi} =
\left(\bar{\mathbf{u}},\bar{\mathbf{d}},\bar{\mathbf{s}}\right)$ which
corresponds to the four dimensional triplet $\bar{\psi}
=\left(\bar{u},\bar{d},\bar{s}\right)$. 

These fields are used to build an
interpolating operator $\mathbf{B}$ with the same spin and flavour structure of
the corresponding 4D local operators $\mathcal{B}(x)$ which we have defined
above. This new operator is coupled with pseudoscalar bilinears of the form
$\bar{\varphi}\gamma_5\psi$. Computations using baryonic two point functions
have in fact shown that the signal obtained with the pseudoscalar bilinears has
much lower noise contributions than the one given by the scalar bilinears.
The coupling of $\mathbf{B}$ with bilinears allows the quenched three
dimensional fermions to propagate in time via the four dimensional ones.

It is
now possible to join together the pieces and define the 3D extended operators
\begin{equation}
\label{eq:3dop}
O(t,\mathbf{\mathbf{w}}) =a^9\sum_{\mathbf{x},\mathbf{y},\mathbf{z}} \mathbf{B}(\mathbf{w})\bar{\varphi}\gamma_5\psi(t,\mathbf{x})\bar{\varphi}\gamma_5\psi(t,\mathbf{y})\bar{\varphi}\gamma_5\psi(t,\mathbf{z}).
\end{equation}
There are as many bilinear operators as there are quarks in the interpolating operator  $\mathbf{B}$, for instance if we were to study mesons there would have been only two bilinears instead of three. Due to the fact that the 3D fermion action is symmetric under $C$,$P$ and $\Gamma$ symmetry\footnote{$\varphi \rightarrow e^{i\alpha \Gamma} \varphi$ and $U_\mu(x) \rightarrow U_\mu(x)$ where $\Gamma=i\gamma_0\gamma_5$} it can be shown that the 3D extended operators are well behaved under renormalisation.\par
From the operator of Eq.~(\ref{eq:3dop}), the smoothed baryonic two point functions 
are constructed  in the usual way from products of quark propagators. For the 3D extended operators, this amounts to replacing
each 4D propagator in the formula for the point sources with the product 
\begin{equation}
\label{3D}
S(t,\mathbf{w},t',\mathbf{w'}) = a^6\sum_{\mathbf{x},\mathbf{x'}}\,S_{3D}(\mathbf{w},\mathbf{x'})\,\gamma_5\,
S_{4D}(t,\mathbf{x'};t',\mathbf{x})\,\gamma_5\, S_{3D}(\mathbf{x};\mathbf{w'})\,.
\end{equation}
It can be shown from dimensional considerations, that the 3D extended
operator's two point function is much more regular at short distances than the
one where local operators are used. Indeed the degree of divergence in time
will be at most logarithmic, while for the local operators it is a polynomial of
degree six. 

\begin{figure}[!tbp]
	\centering
	\includegraphics[width=0.45\textwidth]{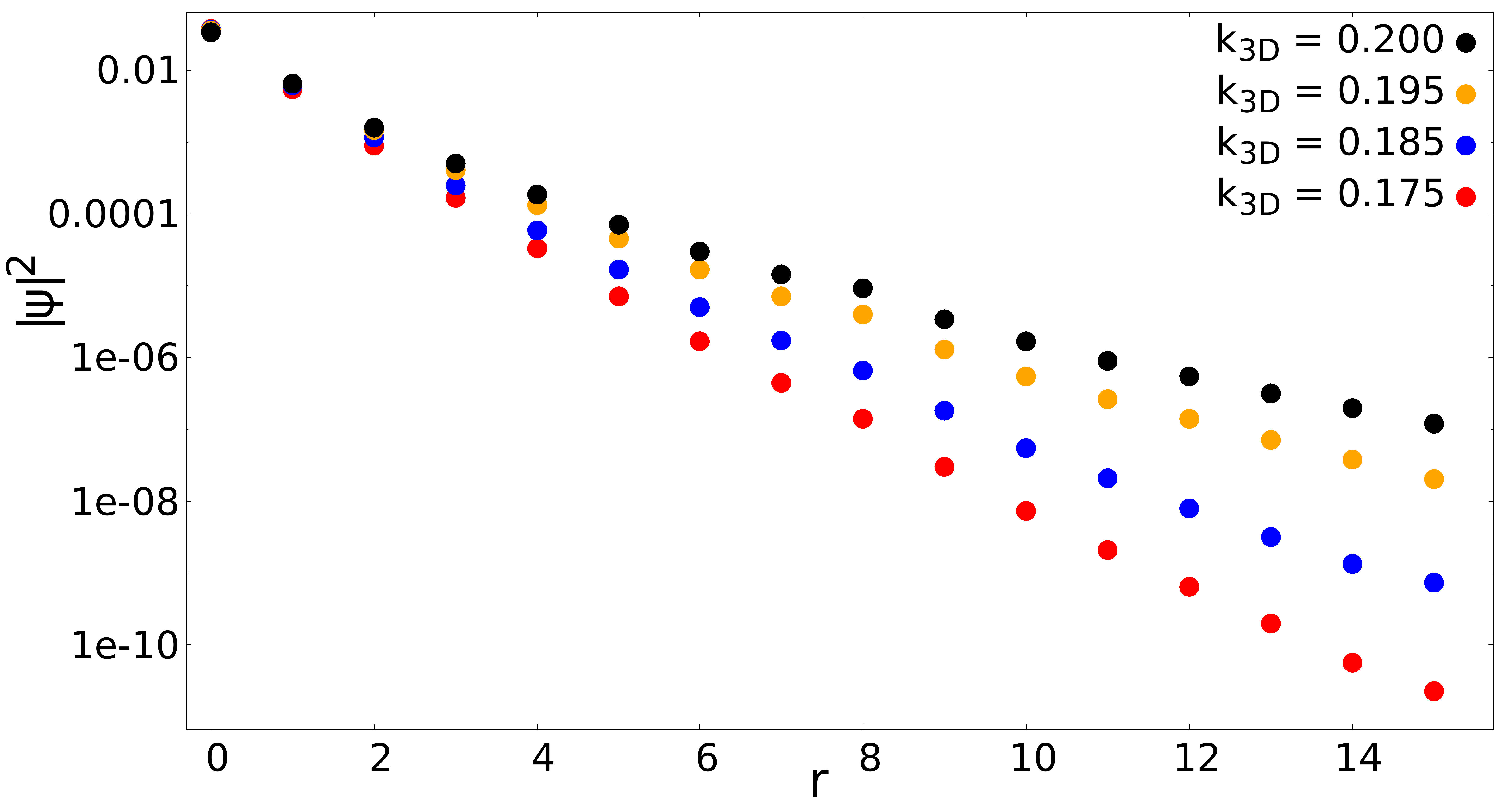}
  \hspace{0.5cm}
	\includegraphics[width=0.45\textwidth]{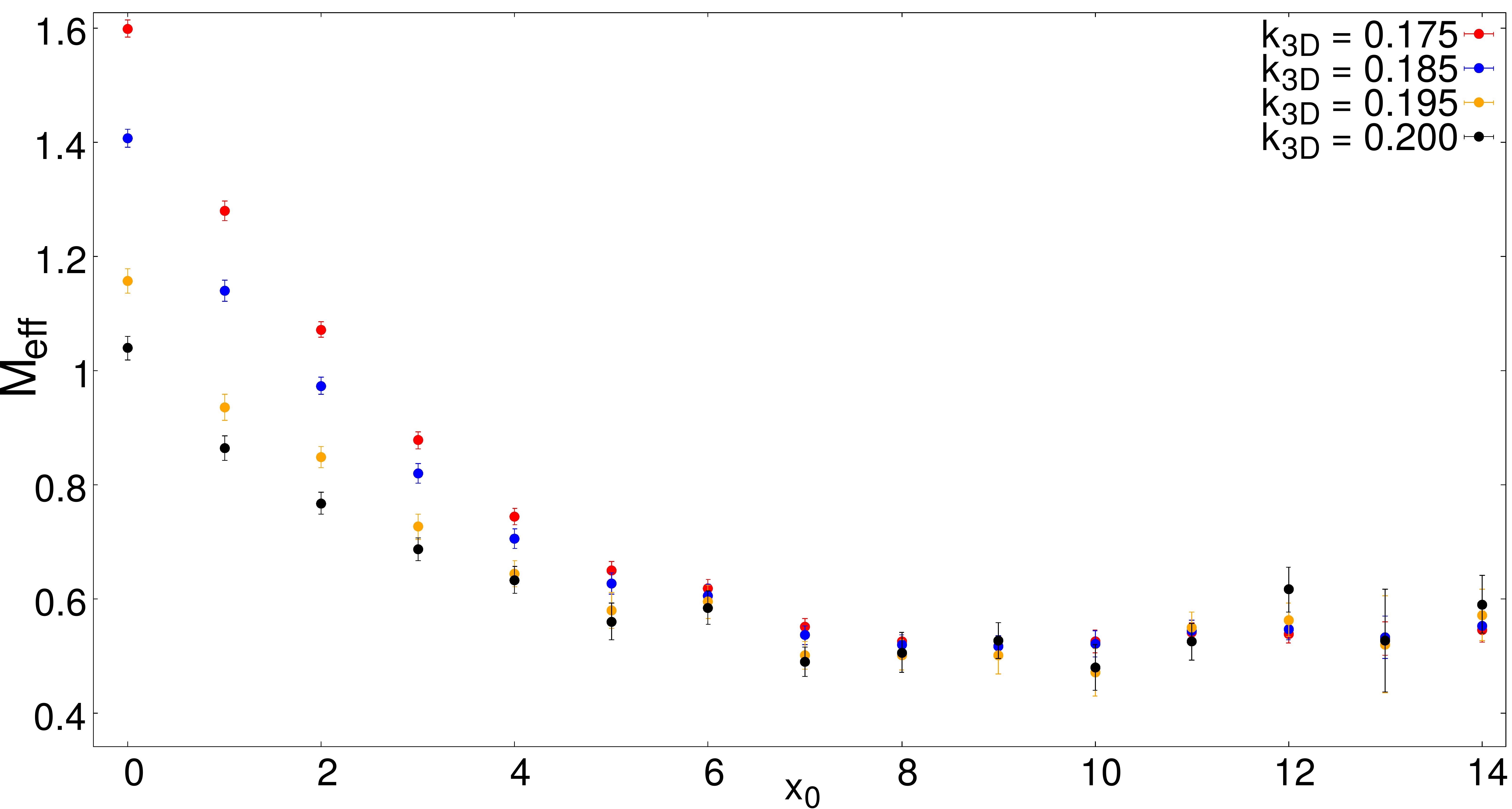}
	\caption{Left: Norm square of the 3D source $\rvert \Psi\rvert^2$ along one axis as a function of distance for the extended operators. For $\kappa_{3D}\rightarrow \kappa^{crit}_{3D}$ the 3D fields are much more spread out. Right:
	Effective mass of the nucleon for various values of $\kappa_{3D}$. The larger $\kappa_{3D}$, the stronger the excited states are suppressed. }
	\label{fig:3Dkappas}
\end{figure}
\section{Numerical tests}

The new operators have been numerically implemented and  compared to point sources and Jacobi smearing. The lattice simulations were performed on
the  CLS $N_f = 2+1$ gauge configurations \cite{Bruno:2014jqa} that adopt
the open boundary conditions in time \cite{Luscher:2011kk}. In the
present exploratory study we have a flavour $SU(3)$ symmetric $96\times32^3$ 
ensemble with
$M_\pi=M_K=420$\,MeV and a  lattice spacing $a = 0.086$\,fm \cite{Bruno:2016plf}.

\subsection{Shape of the source}

For this first test, we use the same 3D mass for all fermion fields,
defining analogously to the 4D case
$m_{3D} = \frac{1}{2}\left( \frac{1}{\kappa_{3D}}-6\right)$. 
The norm of the wave function as a function of the 
separation from the source is shown  for select values
of $\kappa_{3D}$ in Fig.~\ref{fig:3Dkappas}. As expected, a smaller 3D mass corresponds to a source which is more spread out,
with the square of the effective mass being proportional to the 3D mass in a wide range.

For our value of the physical parameters,  a  critical value of $\kappa_{3D}$
can be found, at which the effective mass of the three-dimensional analogue to
the pion goes to zero. A linear fit to this behavior leads to an estimate of
$\kappa_{3D}^{crit} = 0.208\pm 0.004$ for this ensemble. For the rest of the
study, we employed  $\kappa_{3D} = 0.185$, which we have chosen to match
standard parameters of Jacobi sources as explained below.

In the right hand plot of Fig.~\ref{fig:3Dkappas}, the effect of the extension of the source
on the nucleon two-point function is shown. The smaller $m_{3D}$, i.e. the wider the source,
the stronger is the suppression of the excited state contribution at small distances.
Nevertheless, the plateau region from which the ground state mass is extracted, 
starts roughly at the same time separation.

\begin{figure}[!tbp]
	\centering
	\includegraphics[width=0.5\linewidth]{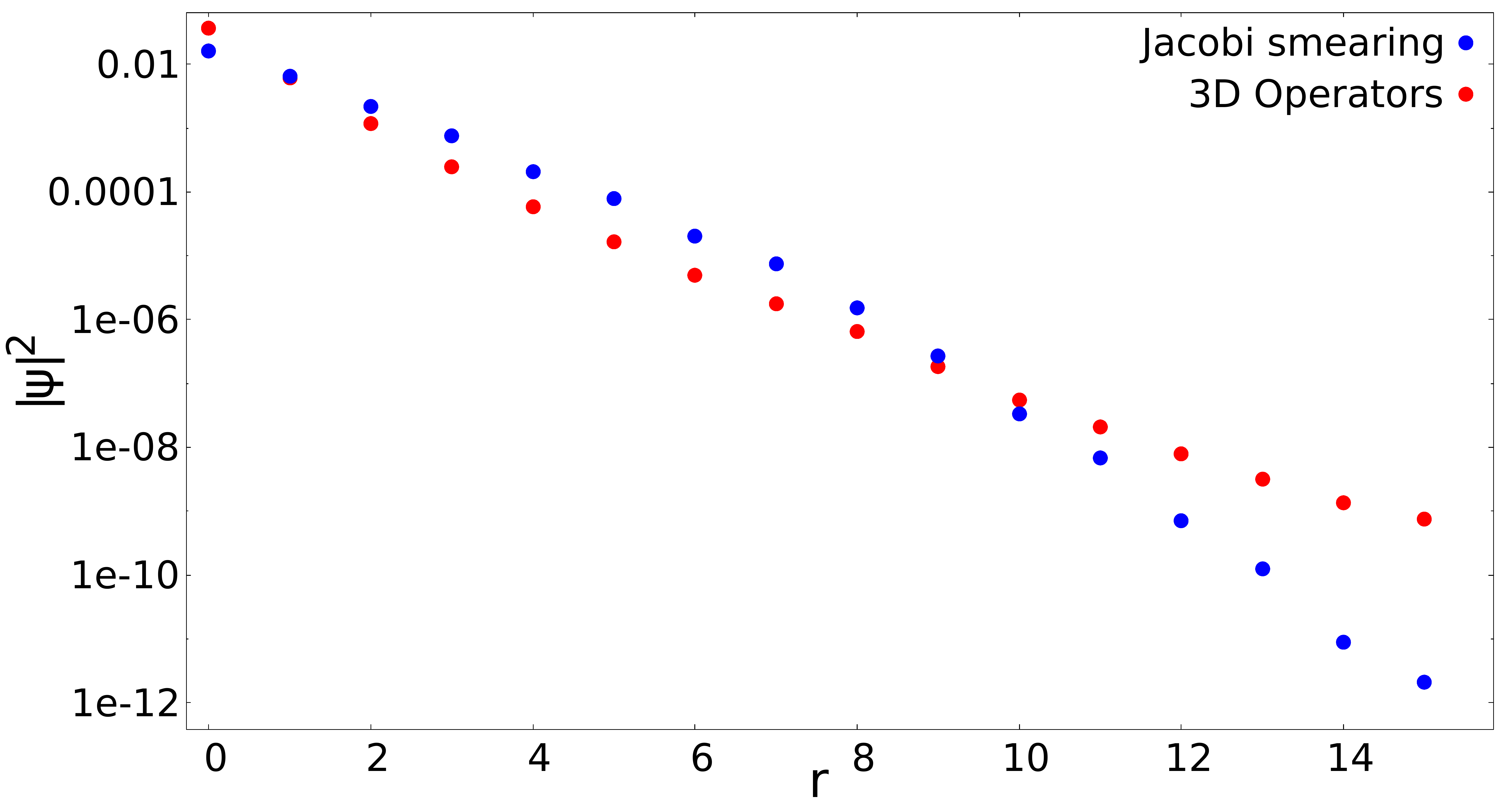}
	\caption{The norm of the 3D source (red) and the Jacobi smeared source (blue) as a function of distance $r$ for the extended operators. The parameters of the Jacobi smearing were tuned to match the average radii $\langle r^2 \rangle$. }
	\label{fig:psisquarekappas2}
\end{figure}

\subsection{Comparison to Jacobi smearing}

\begin{figure}[!tbp]
	\centering
  \subfloat[$\hat{p}$ from eq.(\ref{proton12})]{\includegraphics[width=0.49\textwidth,height=\textheight,keepaspectratio]{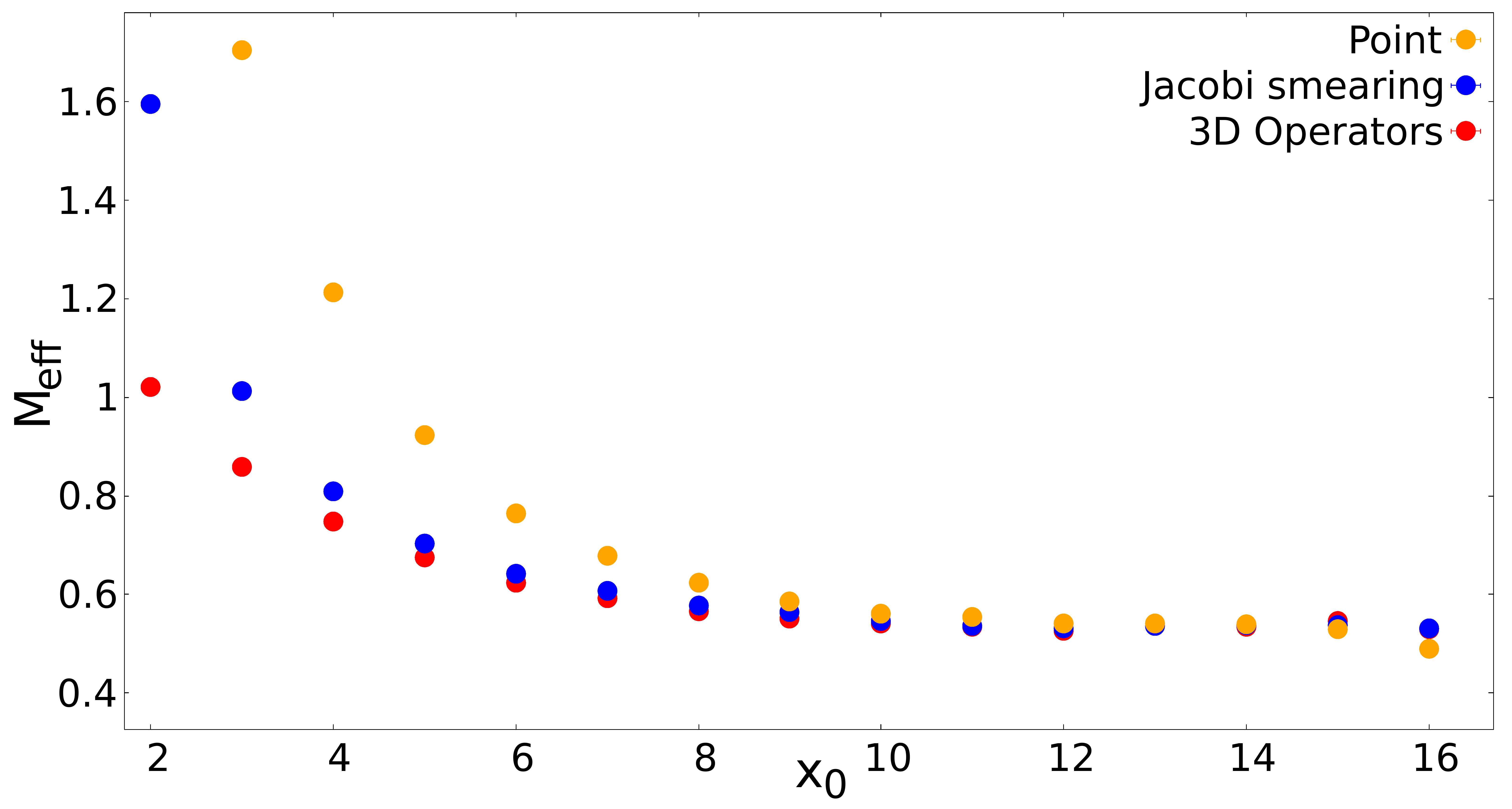}}
	\hfill
	\subfloat[$\tilde{p}$ from eq.(\ref{proton12})]{\includegraphics[width=0.49\textwidth]{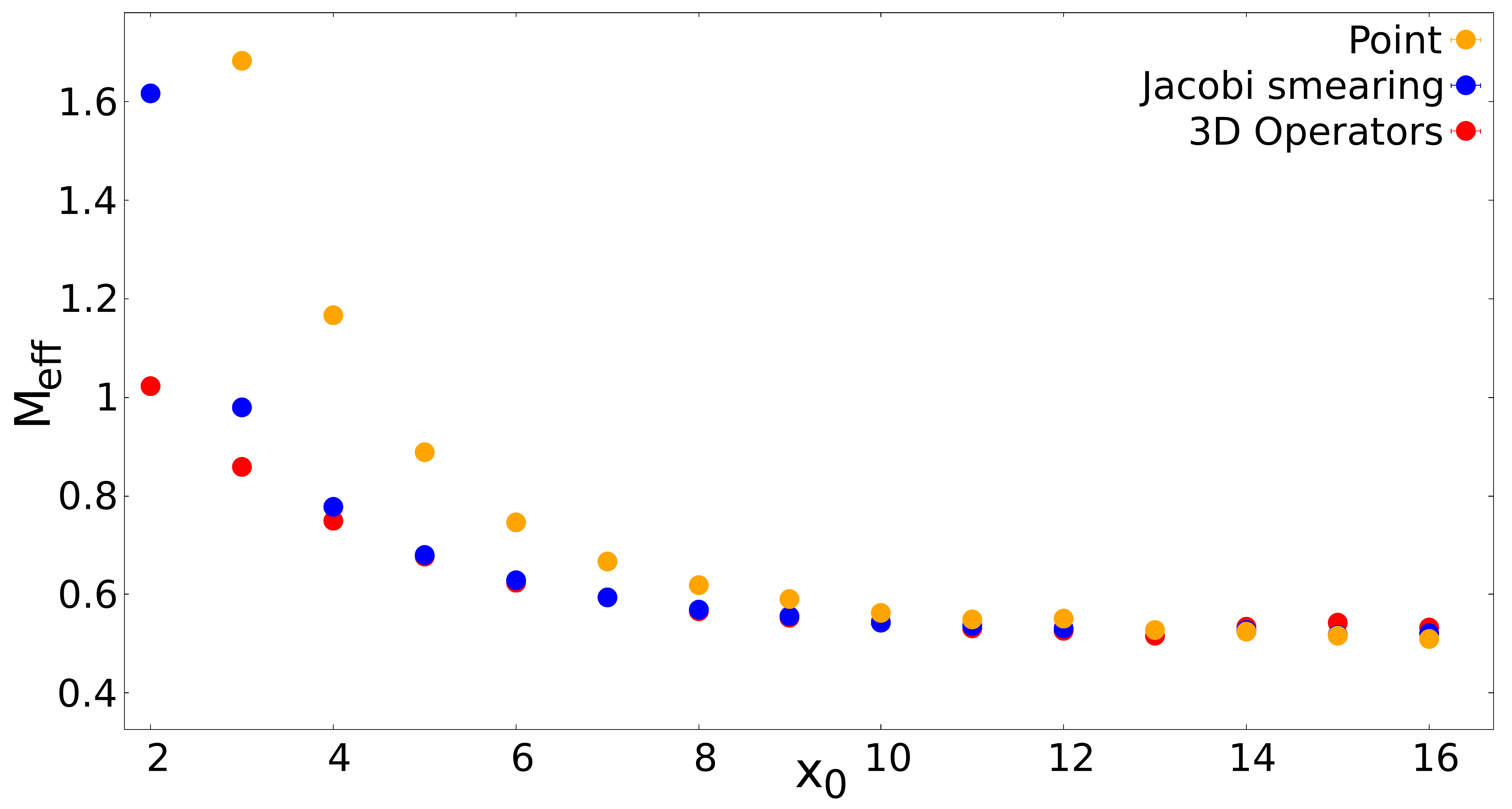}}
	\hfill
	\subfloat[$p$ from eq.(\ref{proton3})]{\includegraphics[width=0.49\textwidth]{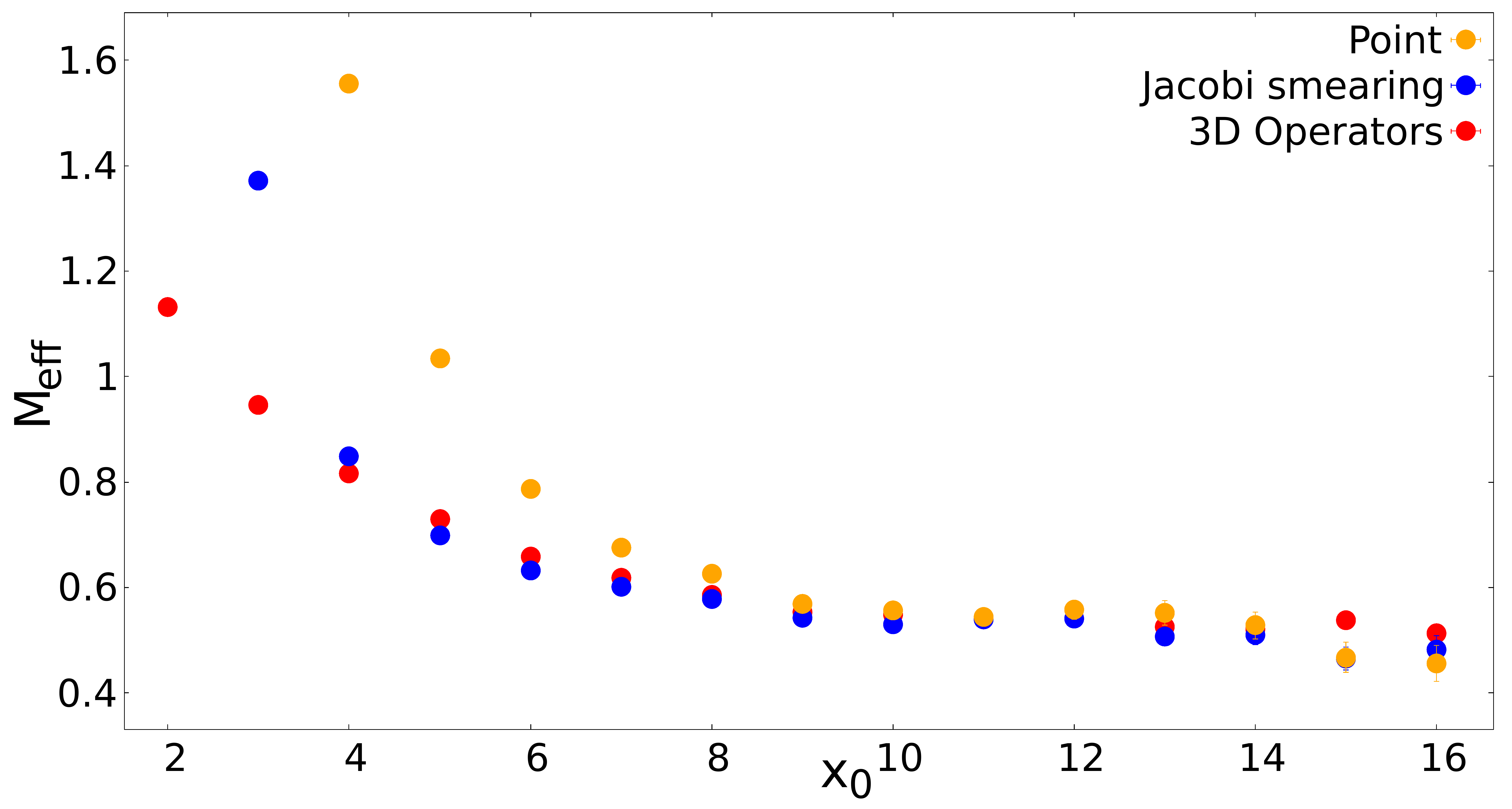}}
  \caption{Comparison between the effective masses of the nucleon
  calculated for the 3D extended operators (red) and the Jacobi smearing (blue) and point sources (orange)
for  the three different nucleon  operators of eqs.(\ref{proton12}) and
(\ref{proton3}). The 3D extended operators have a better suppression of the
contributions from the excited states and also the short distance behaviour is
much more regular. While similar from about $x_0=4a$ on, the  Jacobi smeared sources
have a stronger coupling to the excited states at smaller distance.}
	\label{figure_n}
\end{figure}

\begin{figure}[!tbp]
	\centering
	\subfloat[$\Omega^{(\frac{1}{2},1)}$ from eq.(\ref{omega12})]{\includegraphics[width=0.49\textwidth,height=\textheight,keepaspectratio]{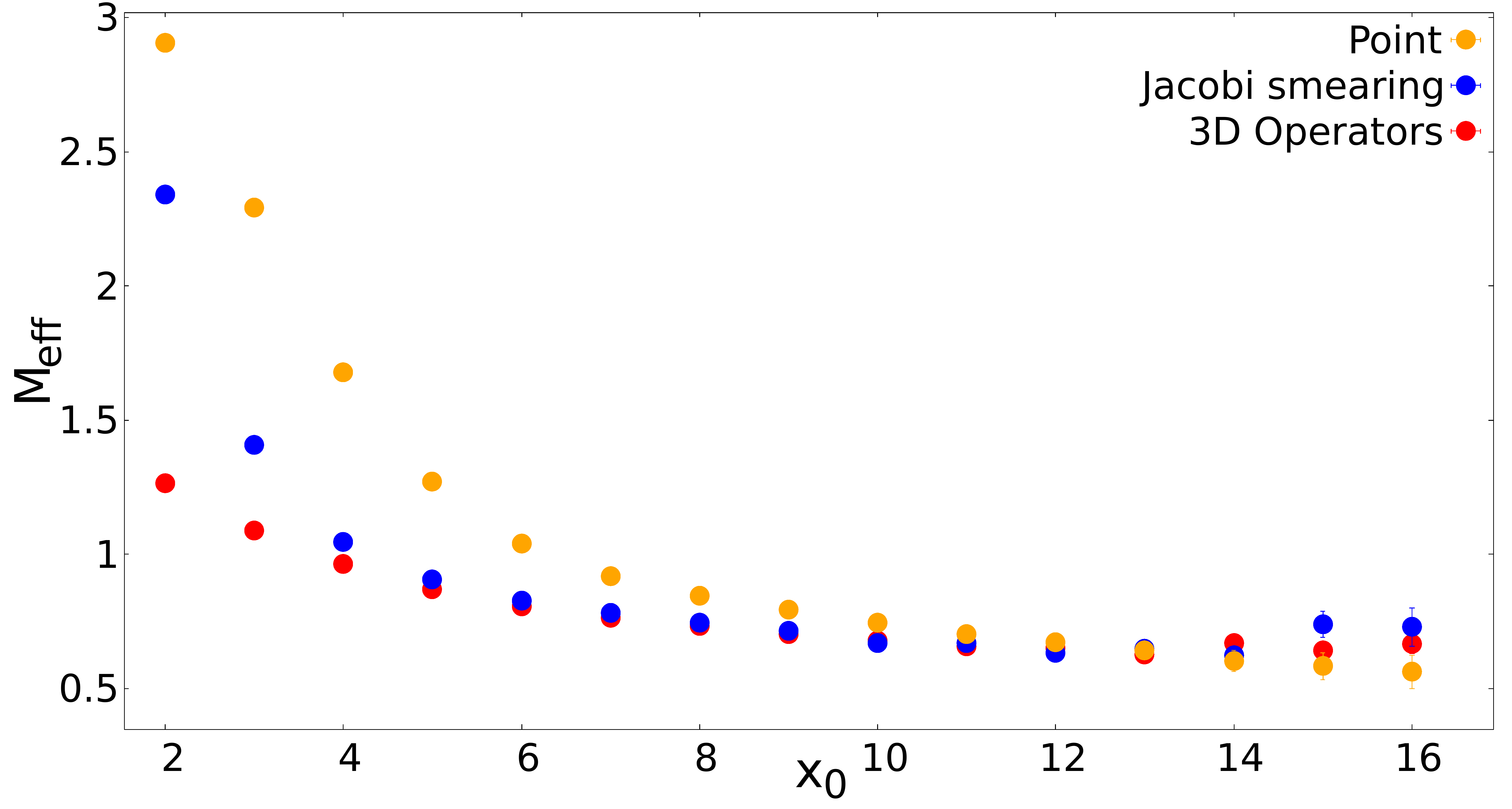}\label{fig:f1a}}
	\hfill
	\subfloat[$\Omega^{(\frac{3}{2},0)}$ from eq.(\ref{omega12})]{\includegraphics[width=0.49\textwidth]{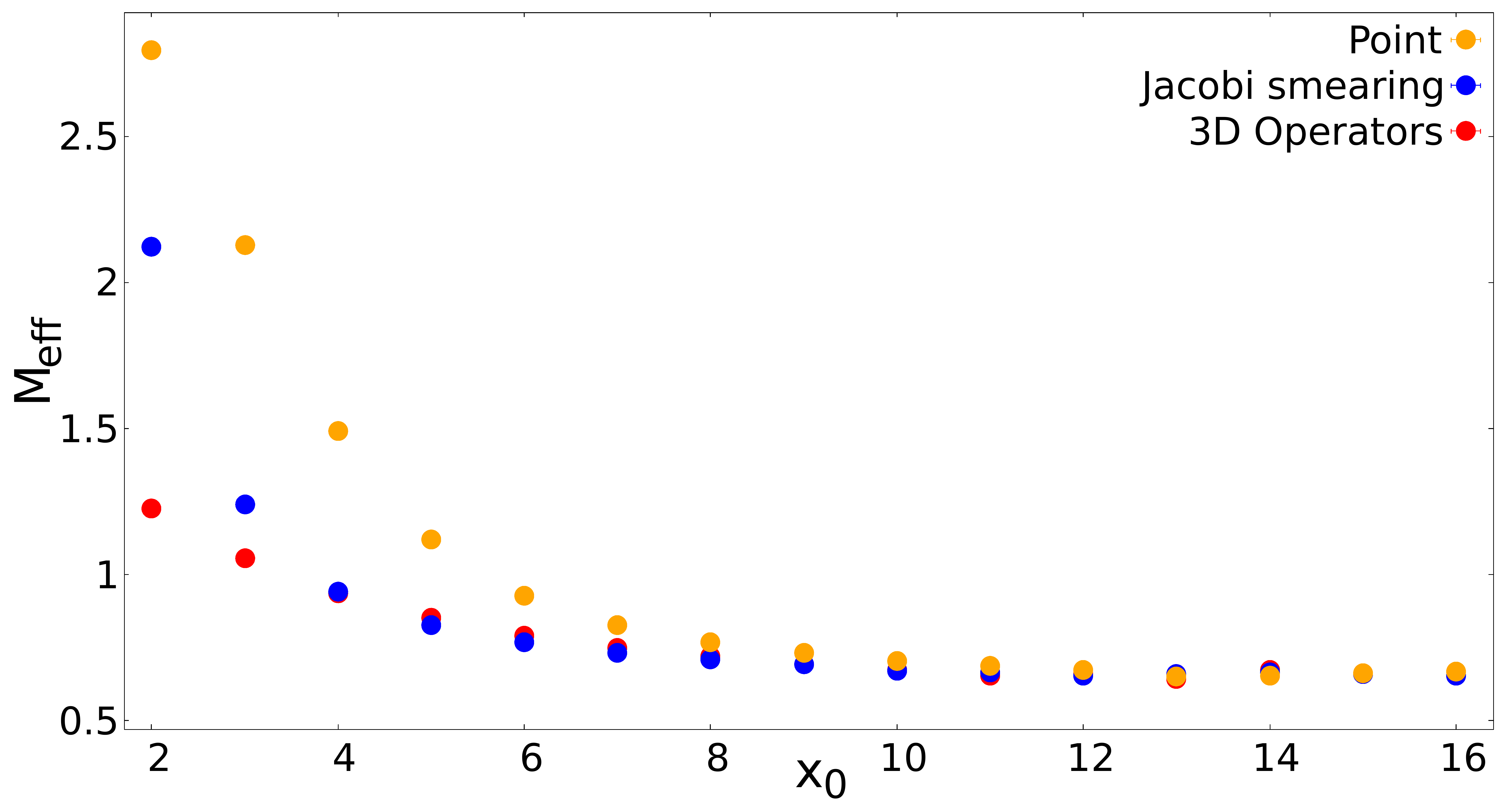}\label{fig:f2c}}
	\caption{The comparison between the effective masses of the Omega calculated for the 3D extended operators (red) and the Jacobi smearing (blue) and point sources (orange). The effective mass is extracted from the two point functions built from the two different Omega operators of eq.(\ref{omega12}). It is possible to draw similar conclusions to the nucleon case.}
		\label{figure_o}
\end{figure}

To test the effectiveness of the 3D extended operator technique relative to the existing ones, we have computed the baryonic two point functions with  standard  point sources and Jacobi smeared sources.
The free parameters of the Jacobi smearing are the number of terms included into the sum $N_{sm}$ and $\kappa_{sm}$ which regulates how strongly the source is spread out \cite{Allton:1993wc}  
\begin{equation}
  \Psi_\mathrm{sm} = \sum_{n = 0}^{N_{sm}} (\kappa_{sm}\Delta)^{n}\  \Psi_\mathrm{pnt} \,.
\end{equation}
The parameters have been chosen to be $N_{sm}  = 50$ and $\kappa_{sm} = 0.21$,
such that the square of the source radius matches to the one of the 3D fermions.
The Jacobi smeared source $\Psi_{sm}$ behaves very differently from
the 3D one as can be seen in Fig.~\ref{fig:psisquarekappas2}. The latter has the typical exponential 
decay in the long distance region, whereas the Jacobi source has a the shape similar to a Gaussian. 
The different shapes of these two source types give the opportunity to construct operators that differ
significantly from one another. Therefore they might be well suited to
construct an enlarged basis to be used with variational methods.

Finally, Figs.~\ref{figure_n} and \ref{figure_o} show the effective masses 
of the nucleon and the Omega, respectively, as a function of source sink separation.  For both baryons, the  3D
extended operators have an excellent short distance behaviour compared to the
other two methods. On the other hand, the plateaux for both, the 3D extended operators and
Jacobi smearing, are  reached at about the same  time  and have very similar statistical noise.
 
\section{Summary}
The 3D fermions introduced in this poster 
are an interesting alternative to the more standard Jacobi smearing
and look promising in increasing the precision in the extraction of the baryon
spectrum. 
In the numerical application, the 3D propagators can be computed with standard 
iterative techniques for sparse systems. This can make them a computationally
efficient choice for the construction of wide sources.

It can be shown that these operators improve the short distance
behaviour of two point functions and that they are well behaved under
renormalisation.  The 3D extended operators therefore allow to have theoretical control
while taking the continuum limit. The disappearance of short distance
divergences might improve the study of excited states, where a
strong signal is needed at short Euclidean times.  However a satisfactory understanding of the  interplay between the short distance behaviour and excited states can only be achieved 
through a scaling study, one of the next steps we are planning. 

The previous properties together with their difference from the Jacobi smeared sources and the freedom given by the
tunable $m_{3D}$ mass parameter make the 3D extended operators a potentially very interesting addition to the basis 
used in the GEVP and therefore promise to help in the extraction of the spectrum.

We thank Martin L\"{u}scher for many discussions on the subject. 

\bibliographystyle{JHEP}
\bibliography{mybibb} 

\end{document}